\documentclass{article}
\usepackage{spconf,amsmath,graphicx}
\usepackage{multirow,booktabs}
\usepackage[normalem]{ulem}
\usepackage{xspace}
\def\onedot{.\@\xspace}
\def\etal{\emph{et al}\onedot}
\useunder{\uline}{\ul}{}
\usepackage[table,x11names]{xcolor}
\newcommand{\mycc}{\cellcolor[HTML]{f2f2f2}}
\definecolor{colone}{RGB}{133, 0, 3}

\definecolor{coltwo}{RGB}{190, 81, 7}

\definecolor{colthree}{RGB}{13,85, 2}

\definecolor{colfour}{RGB}{8, 0, 135}

\title{RECAP: Retrieval-Augmented Audio Captioning}
\name{Sreyan Ghosh, Sonal Kumar, Chandra Kiran Reddy Evuru, Ramani Duraiswami, Dinesh Manocha}
\address{University of Maryland, College Park, USA}
\begin{document}
%
\maketitle
\begin{abstract}
We present \textbf{RECAP} (\textbf{RE}trieval-Augmented Audio \textbf{CAP}-tioning), a novel and effective audio captioning system that generates captions conditioned on an input audio and other captions similar to the audio retrieved from a datastore. Additionally, our proposed method can transfer to any domain without the need for any additional fine-tuning. To generate a caption for an audio sample, we leverage an audio-text model CLAP~\cite{wu2023large} to retrieve captions similar to it from a replaceable datastore, which are then used to construct a prompt. Next, we feed this prompt to a GPT-2 decoder and introduce cross-attention layers between the CLAP encoder and GPT-2 to condition the audio for caption generation. Experiments on two benchmark datasets, Clotho and AudioCaps, show that RECAP achieves competitive performance in in-domain settings and significant improvements in out-of-domain settings. Additionally, due to its capability to exploit a large text-captions-only datastore in a \textit{training-free} fashion, RECAP shows unique capabilities of captioning novel audio events never seen during training and compositional audios with multiple events. To promote research in this space, we also release 150,000+ new weakly labeled captions for AudioSet, AudioCaps, and Clotho\footnote{https://github.com/Sreyan88/RECAP}.
\end{abstract}





%
\section{Introduction}
\label{sec:intro}

Audio captioning is the fundamental task of describing the contents of an audio sample using natural language. Compared to Automatic Speech Recognition (ASR), which transcribes human speech, audio captioning focuses on describing distinct environmental sounds in the input audio \cite{kim2019audiocaps,drossos2020clotho}. By bridging the gap between text and audio modalities, audio captioning has found various applications in real-world use cases like environment monitoring, gaming, etc. \cite{koepke2022audio}.

\begin{figure}
    \centering
    \includegraphics[width=\columnwidth]{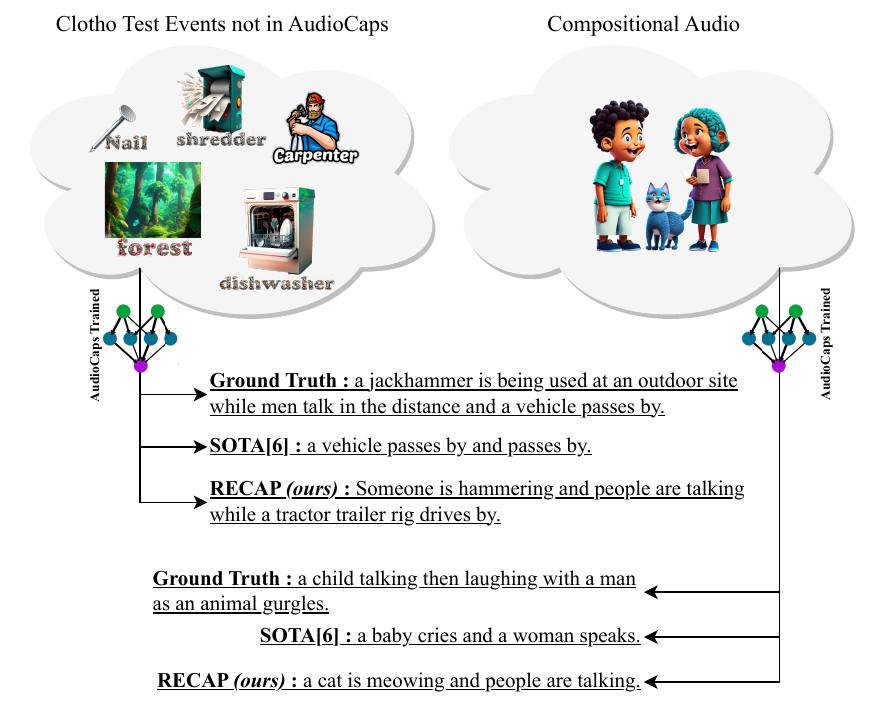}
    \caption{\small We propose \textbf{RECAP}, a retrieval-augmented audio captioning model. RECAP can caption novel concepts never before seen in training and improves the captioning of audio with multiple events.}
    \label{fig:enter-label}
\end{figure}

In the past, most audio captioning models employed an encoder-decoder architecture using an off-the-shelf pre-trained audio encoder and a language decoder \cite{gontier2021automated,kim2023prefix}. The audio encoder generates an audio embedding sequence that is used to condition the language decoder for caption generation. However, most of these systems do not perform well on cross-domain settings (trained on one domain and tested on the other), and every use case might need separate training. We hypothesize that the primary reason behind this phenomenon is the shift of occurrence of unique audio events with a domain shift. For example, the AudioCaps benchmark dataset \cite{kim2019audiocaps} has several audio concepts (e.g., the sound of jazz or an interview) that Clotho, another benchmark dataset, does not. This is also representative of real-world scenarios where not only do audio concepts change from one domain to another (e.g., environmental sounds in a city versus a forest), but new audio concepts also keep emerging within a domain (e.g., new versions of an online game). 
\vspace{1mm}

\begin{figure*}[t!]
  \centering
\includegraphics[width=2\columnwidth]{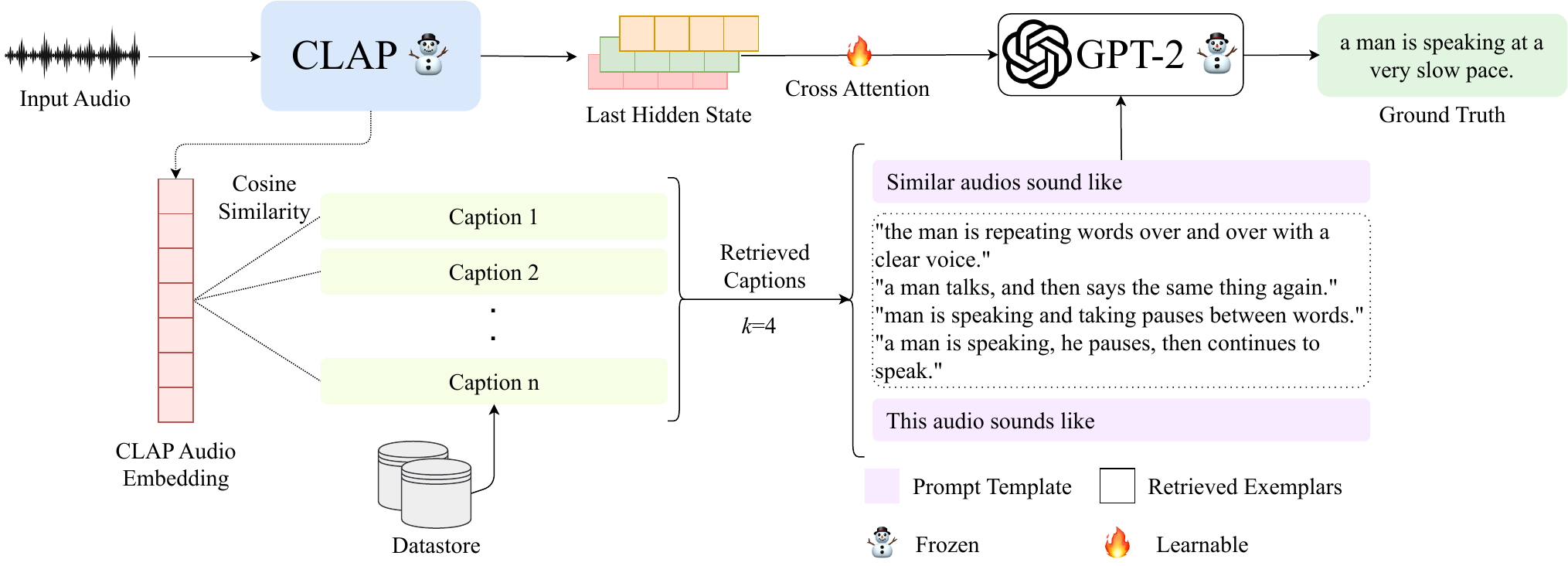}
\vspace{-1em}
  \caption{\small Illustration of \textbf{RECAP}. RECAP fine-tunes a GPT-2 LM conditioned on audio representations from the last hidden state of CLAP~\cite{wu2023large} and a text prompt. The text prompt is constructed using captions most similar to the audio, retrieved from a datastore using CLAP.}
  \label{fig:model}
\end{figure*}

{\noindent \textbf{Main Contributions.}} We propose RECAP, \textbf{RE}trieval-Augmented Audio \textbf{CAP}tioning, a simple and scalable solution to the aforementioned problems of domain shifts. Similar to other audio captioning systems in literature \cite{gontier2021automated, kim2023prefix,koizumi2020audio}, RECAP is built on an audio encoder and a language decoder (GPT-2 in our setting). However, we introduce three novel changes: (1) Instead of employing an audio encoder pre-trained only on audio, we use CLAP \cite{wu2023large} as our audio encoder. CLAP is pre-trained on audio-text pairs to learn the correspondence between audio and text by projecting them into a shared latent space. Thus, CLAP hidden state representations are better suited for captioning due to their enhanced linguistic comprehension. (2) We condition the audio for caption generation by introducing new cross-attention layers between CLAP and the GPT-2. (3) Finally, beyond just conditioning audio, we also condition a custom-constructed prompt for training and inference. We construct the prompt using the top-\textit{k} captions most similar to the audio from a datastore retrieved using CLAP. We provide more details in Section \ref{subsec:recap}. RECAP builds on retrieval-augmented generation (RAG) \cite{lewis2020retrieval}, which offers multiple advantages discussed further in Section \ref{sec:methodology}. RECAP is lightweight, fast to train (as we only optimize the cross-attention layers), and can exploit any large text-caption-only datastore in a \textit{training-free} fashion. We evaluate RECAP on two benchmark datasets, Clotho \cite{drossos2020clotho} and AudioCaps \cite{kim2019audiocaps}, and show that while being competitive to the state-of-the-art in in-domain settings, RECAP outperforms all baselines in out-of-domain settings by a large margin. Additionally, RECAP can effectively caption novel audio events never seen during training and can better generate captions for compositional audios with multiple audio events.

\section{Related Work}
\label{sec:related_work}

{\noindent \textbf{Automated Audio Captioning.}} Current work in audio captioning primarily employs encoder-decoder models where a caption is generated by an autoregressive language decoder conditioned on representations obtained from an audio encoder \cite{gontier2021automated, kim2023prefix,koizumi2020audio}. The language decoder employed is either pre-trained on web-scale data \cite{gontier2021automated, kim2023prefix,koizumi2020audio} or learned from scratch \cite{eren2020audio,xu2021investigating} during fine-tuning. The work closest to ours is \cite{koizumi2020audio}, where the authors condition a GPT-2 on prompts constructed using retrieved captions. However, the key difference between our work and theirs is that we require only a text-caption-only datastore for RECAP, whereas their system requires both audio and text pairs. We also introduce additional cross-attention layers for audio conditioning. \textit{Kim et. al}~\cite{kim2023prefix}, the current state-of-the-art system, proposed prefix tuning for audio captioning where the authors feed a prefix or a fixed-size embedding sequence to GPT-2 for audio captioning. Other works include synthetic data augmentation techniques \cite{wu2023_t6a,kadlvcik2023whisper}, and training tricks to improve learning on the source training data \cite{yan2023_t6a,lee2023_t6a}.
\vspace{1mm}

{\noindent \textbf{Retrieval-augmented Generation.}} The core idea of retrieval-augmented generation (RAG) is to condition generation on additional data retrieved from an external datastore \cite{lewis2020retrieval}. RAG has been shown to benefit knowledge-intensive NLP tasks like open-domain question-answering on datasets that require world knowledge and advanced reasoning capabilities \cite{berant2013semantic,kwiatkowski2019natural}. RAG has also proven to be extremely effective in various computer vision tasks, including image captioning \cite{ramos2023retrieval,Ramos_2023_CVPR}. We argue that audio captioning, especially in out-of-domain scenarios, is a knowledge-intensive task as it requires the model to caption novel audio concepts never seen during training, and can benefit from RAG.



\begin{table*}[t]
\footnotesize
\centering
     \caption{\small Evaluation on Clotho. Each method is trained on three different settings and tested on the AudioCaps dataset. For evaluation, we use a datastore that has captions from the training set ($\mathcal{DS}$), AudioCaps ($\mathcal{DS}_{caps}$), or a large external dataset ($\mathcal{DS}_{large}$).}
    \resizebox{0.9\linewidth}{!}{
    \begin{tabular}{ccccccccccc}
    \toprule
    Training set & Method&BLEU$_1$&BLEU$_2$&BLEU$_3$&BLEU$_4$&METEOR&ROUGE$_L$&CIDEr&SPICE&SPIDEr\\
    \hline
    \multirow{8}{*}{(1) Clotho} & Mei~\etal~\cite{mei2021audio} & 0.527 & 0.327 & 0.211 & 0.131 & 0.158 & 0.356 & 0.320 & 0.105 & 0.213\\ 
    & Gontier~\etal~\cite{gontier2021automated} & 0.506 & 0.318 & 0.210 & 0.134 & 0.148 & 0.338 & 0.278 & 0.092 & 0.185\\ 
    & Chen~\etal~\cite{chen2020audio} & 0.534 & 0.343 & 0.230 & 0.151 & 0.160 & 0.356 & 0.346 & 0.108 & 0.227 \\
    & Xu~\etal~\cite{xu2021investigating} & 0.556 & 0.363 & 0.242 & 0.159 & 0.169 & 0.368 & 0.377 & 0.115 & {\ul0.246} \\
    & Koh~\etal~\cite{koh2022automated} & 0.551 & 0.369 & 0.252 & \textbf{0.168} & 0.165 & 0.373 & 0.380 & 0.111 & {\ul0.246} \\
    & Kim~\etal~\cite{kim2023prefix} & 0.560 & 0.376 & {\ul0.253} & 0.160 & 0.170 & 0.378 & 0.392 & 0.118 & \textbf{0.255} \\
    &\mycc RECAP (w/ $\mathcal{DS}$) &\mycc {\ul 0.563} &\mycc {\ul0.381} &\mycc \textbf{0.257} &\mycc 0.165 &\mycc \textbf{0.179} &\mycc {\ul0.383} &\mycc {\ul0.398} &\mycc {\ul0.122} &\mycc 0.214 \\
    &\mycc RECAP (w/ $\mathcal{DS}_{large}$) &\mycc \textbf{0.582} &\mycc \textbf{0.384} &\mycc \textbf{0.257} &\mycc {\ul 0.166} &\mycc {\ul0.177} &\mycc \textbf{0.395}&\mycc \textbf{0.411} &\mycc \textbf{0.125} &\mycc 0.224 \\
    
    \hline
    \multirow{7}{*}{(2) AudioCaps} & Mei~\etal~\cite{mei2021audio} & 0.294 & 0.146 & 0.080 & 0.043 & 0.096 & 0.239 & 0.117 & 0.050 & 0.084 \\
    & Gontier~\etal~\cite{gontier2021automated} & {0.309} & {0.146} & 0.071 & 0.034 & 0.098 & 0.233 & 0.112 & 0.046 & 0.079\\ 
    & Chen~\etal~\cite{chen2020audio} & 0.226 & 0.114 & 0.065 & 0.039 & 0.086 & 0.228 & 0.109 & 0.042 & 0.076 \\
    
    & Kim~\etal~\cite{kim2023prefix} & 0.342 & 0.195 & 0.115 & 0.065 & 0.112 & 0.276 & 0.192 & 0.074 & 0.133 \\
    &\mycc RECAP (w/ $\mathcal{DS}_{caps}$) &\mycc 0.339 &\mycc 0.193 &\mycc 0.109 &\mycc 0.068 &\mycc 0.110 &\mycc 0.276 &\mycc 0.195 &\mycc 0.084 &\mycc 0.137 \\
    &\mycc RECAP (w/ $\mathcal{DS}$) &\mycc {\ul0.515} &\mycc {\ul0.349} &\mycc {\ul0.210} &\mycc {\ul0.143} &\mycc {\ul0.155} &\mycc \textbf{0.328}&\mycc \textbf{0.332} &\mycc {\ul0.988} &\mycc {\ul0.201} \\
    &\mycc RECAP (w/ $\mathcal{DS}_{large}$) &\mycc \textbf{0.519} &\mycc \textbf{0.355} &\mycc \textbf{0.216} &\mycc \textbf{0.149} &\mycc \textbf{0.157} &\mycc {\ul0.324} &\mycc {\ul0.331} &\mycc \textbf{1.004} &\mycc \textbf{0.209} \\
    \hline 
    
    \multirow{6}{*}{\begin{tabular}[c]{@{}c@{}}(3) Clotho \&\\ AudioCaps\end{tabular}} & Mei~\etal~\cite{mei2021audio} & 0.516 & 0.318 & 0.204 & 0.127 & 0.157 & 0.351 & 0.313 & 0.105 & 0.209 \\
    & Gontier~\etal~\cite{gontier2021automated} & 0.461 & 0.282 & 0.182 & 0.117 & 0.136 & 0.318 & 0.251 & 0.083 & 0.167 \\
    & Chen~\etal~\cite{chen2020audio} & 0.516 & 0.325 & 0.215 & 0.141 & 0.153 & 0.350 & 0.314 & 0.102 & 0.208 \\
    & Kim~\etal~\cite{kim2023prefix} & 0.539 & 0.346 & {\ul0.227} & 0.142 & 0.159 & 0.366 & 0.319 & {\ul0.111} & 0.215 \\
    &\mycc RECAP (w/ $\mathcal{DS}$) &\mycc {\ul0.547} &\mycc \textbf{0.361} &\mycc \textbf{0.238} &\mycc {\ul0.149} &\mycc \textbf{0.167} &\mycc {\ul0.379} &\mycc {\ul0.322} &\mycc \textbf{0.116} &\mycc \textbf{0.222} \\
    &\mycc RECAP (w/ $\mathcal{DS}_{large}$) &\mycc \textbf{0.549} &\mycc {\ul0.360} &\mycc \textbf{0.238} &\mycc \textbf{0.150} &\mycc {\ul0.166} &\mycc \textbf{0.381}&\mycc \textbf{0.323} &\mycc \textbf{0.116} &\mycc {\ul0.221} \\
    \bottomrule
    \end{tabular}
     }
    \vspace{-3mm}
\label{table:clotho}
\end{table*}

\section{Methodology}
\label{sec:methodology}


{\noindent \textbf{Problem Formulation.}} Given a dataset $\mathcal{D}$ with audio-text pairs ($\mathcal{A}$,$\mathcal{T}$), where each text caption $t_i \in \mathcal{T}$ corresponding to an audio $a_i \in \mathcal{A}$ describes the content or events of the audio, we aim to train a model $\theta$ to generate $t_i$ from $a_i$. Different from other audio captioning systems, we also assume that the model has access to a datastore $\mathcal{DS}$ with text captions during inference. These captions come from the training set of $\mathcal{D}$ or external sources but have no overlap with the validation or test sets of $\mathcal{D}$.

\subsection{RECAP}
\label{subsec:recap}

{\noindent \textbf{Overall Architecture.}} The overall architecture of RECAP is quite simple and lightweight. RECAP employs CLAP as the audio encoder and GPT-2 as the auto-regressive language decoder. To generate the caption, the language decoder conditions on the output of the audio encoder and an individually crafted prompt for each audio. We discuss how we construct the prompt in the next subsection. 

\begin{table*}[t!]
\footnotesize
\centering
     \caption{\small Evaluation on AudioCaps. Each method is trained on three different settings and tested on the AudioCaps dataset. For evaluation, we use a datastore that has captions from the training set ($\mathcal{DS}$), Clotho ($\mathcal{DS}_{clotho}$), or a large external dataset ($\mathcal{DS}_{large}$).}
    \resizebox{0.9\linewidth}{!}{
    \begin{tabular}{ccccccccccc}
    \toprule
    Training set & Method&BLEU$_1$&BLEU$_2$&BLEU$_3$&BLEU$_4$&METEOR&ROUGE$_L$&CIDEr&SPICE&SPIDEr\\
    \hline
    \multirow{7}{*}{(1) AudioCaps} & Mei~\etal~\cite{mei2021audio} & 0.647 & 0.488 & 0.356 & 0.252 & 0.222 & 0.468 & 0.679 & 0.160 & 0.420 \\
    & Gontier~\etal~\cite{gontier2021automated} & 0.699 & 0.523 & 0.380 & 0.266 & 0.241 & 0.493 & \textbf{0.753} & 0.176 & 0.465 \\
    & Chen~\etal~\cite{chen2020audio} & 0.550 & 0.385 & 0.264 & 0.178 & 0.173 & 0.390 & 0.443 & 0.117 & 0.280 \\
    & Eren~\etal~\cite{eren2020audio} & 0.710 & 0.490 & 0.380 & 0.230 & \textbf{0.290} & \textbf{0.590} & 0.750 & - & - \\
    & Kim~\etal~\cite{kim2023prefix} & 0.713 & 0.552 & {\ul0.421} & 0.309 & 0.240 & 0.503 & 0.733 & 0.177 & 0.455 \\
    &\mycc RECAP (w/ $\mathcal{DS}$) &\mycc {\ul0.721}&\mycc \textbf{0.559} &\mycc \textbf{0.428} &\mycc \textbf{0.316} &\mycc 0.252 &\mycc 0.521 &\mycc 0.750 &\mycc {\ul0.183} &\mycc \textbf{0.472} \\
    &\mycc RECAP (w/ $\mathcal{DS}_{large}$) &\mycc \textbf{0.722}&\mycc {\ul0.557} &\mycc \textbf{0.428} &\mycc {\ul0.313} &\mycc {\ul0.256} &\mycc {\ul0.525} &\mycc {\ul0.751} &\mycc \textbf{0.186} &\mycc {\ul0.471} \\
    \hline 
    
    \multirow{7}{*}{(2) Clotho} & Mei~\etal~\cite{mei2021audio} & 0.415 & 0.219 & 0.121 & 0.063 & {0.134} & 0.303 & 0.149 & 0.066 & 0.107 \\
    & Gontier~\etal~\cite{gontier2021automated} & 0.425 & 0.223 & 0.124 & 0.061 & 0.128 & 0.298 & 0.147 & 0.060 & 0.104 \\
    & Chen~\etal~\cite{chen2020audio} & 0.365 & 0.170 & 0.091 & 0.048 & 0.110 & 0.273 & 0.083 & 0.049 & 0.066 \\
    & Kim~\etal~\cite{kim2023prefix} & 0.449 & 0.266 & 0.157 & 0.084 & 0.144 & {\ul0.330} & 0.211 & 0.083 &0.147 \\
    
    &\mycc RECAP (w/ $\mathcal{DS}_{clotho}$)  &\mycc 0.427 &\mycc 0.224 &\mycc 0.148 &\mycc 0.065 &\mycc 0.112 &\mycc 0.281 &\mycc 0.191 &\mycc 0.078 &\mycc 0.136 \\
    &\mycc RECAP (w/ $\mathcal{DS}$) &\mycc  {\ul0.501} &\mycc \textbf{0.326} &\mycc \textbf{0.211} &\mycc {\ul0.104} &\mycc {\ul 0.164} &\mycc \textbf{0.357} &\mycc {\ul0.359} &\mycc \textbf{0.116} &\mycc{\ul0.198} \\
    &\mycc RECAP (w/ $\mathcal{DS}_{large}$) &\mycc \textbf{0.507} &\mycc {\ul0.321} &\mycc {\ul0.206} &\mycc \textbf{0.108} &\mycc \textbf{0.169} &\mycc \textbf{0.357} &\mycc \textbf{0.362} &\mycc {\ul0.111} &\mycc\textbf{0.204} \\
    \hline
    
    \multirow{6}{*}{\begin{tabular}[c]{@{}c@{}}(3) Clotho \&\\ AudioCaps\end{tabular}} & Mei~\etal~\cite{mei2021audio} & 0.682 & 0.507 & 0.369 & 0.266 & 0.238 & 0.488 & 0.701 & 0.166 & 0.434 \\
    & Gontier~\etal~\cite{gontier2021automated} & 0.635 & 0.461 & 0.322 & 0.219 & 0.208 & 0.450 & 0.612 & 0.153 & 0.383 \\
    & Chen~\etal~\cite{chen2020audio} & 0.489 & 0.292 & 0.178 & 0.106 & 0.152 & 0.346 & 0.265 & 0.093 & 0.179 \\
    & Kim~\etal~\cite{kim2023prefix} & 0.708 & 0.547 & 0.402 & 0.283 & 0.238 & {\ul0.499} & 0.710 & 0.167&{\ul 0.438} \\
    &\mycc RECAP (w/ $\mathcal{DS}$) &\mycc \textbf{0.728} &\mycc \textbf{0.563} &\mycc \textbf{0.425} &\mycc {\ul0.317} &\mycc {\ul0.252} &\mycc \textbf{0.529} &\mycc \textbf{0.764} &\mycc {\ul0.187} &\mycc \textbf{0.469} \\
    & \mycc RECAP (w/ $\mathcal{DS}_{large}$) &\mycc {\ul0.725} &\mycc {\ul0.561} &\mycc {\ul0.424} &\mycc \textbf{0.319} &\mycc \textbf{0.256} &\mycc \textbf{0.529} &\mycc {\ul0.761} &\mycc \textbf{0.190}&\mycc \textbf{0.469} \\
    \bottomrule
    \end{tabular}
    }

    \vspace{-3mm}
\label{table:audiocaps}
\end{table*}

For audio conditioning, we first pass the audio samples through the CLAP audio encoder and extract the last hidden state $A \in n \times d$, where $n$ is the sequence length and $d$ is the embedding dimension. This embedding is extracted from the penultimate layer of the CLAP audio encoder right before the final projection. As the audio embeddings and decoder operate on different vector spaces, we connect them through randomly initialized cross-attention modules as each decoder layer. To train the RECAP, we freeze both GPT-2 and the CLAP and only train the cross-attention layers, which reduces overall compute requirements and time for training and retains the expressivity and generalization capabilities of GPT-2. RECAP performs well even after training only 5.4$\%$ of total parameters because, like other retrieval-augmented models \cite{lewis2020retrieval,izacard2022few,li2022survey}, RECAP does not need all information to be stored in its weights as it has access to external knowledge from a datastore of text. Additionally, CLAP generates an audio embedding that correlates well with its corresponding textual description, thus further lowering training time due to its superior understanding of the audio content.
\vspace{1mm}

{\noindent \textbf{Constructing prompts with Retrieved Captions.}} Instead of just conditioning audio features for captioning, RECAP is also conditioned on a prompt, individually crafted for each audio during training and inference. To construct this prompt, RECAP exploits CLAP text and audio encoders \cite{wu2023large}, to retrieve top-\textit{k} captions similar to an audio from a datastore. CLAP encodes audio and text to a shared vector space and has outperformed all prior models on audio-to-text and text-to-audio retrieval, thus making it most suitable for our task. Specifically, for retrieval, we calculate the cosine similarity between the embeddings of the current audio $a_i$ and all the text captions in the datastore $\mathcal{DS}$ and just choose the captions with the highest similarity. Once we have retrieved the top-\textit{k} similar captions, we construct a prompt in the following manner: \textit{``Audios similar to this audio sounds like: caption 1, caption 2, $\cdots$ caption k. This audio sounds like:''}. For retrieval, we naturally ignore the original caption $t_i$ corresponding to $a_i$. RECAP is then trained using the generic cross-entropy loss between the tokens for the predicted caption $\hat{t_i}$ and the ground-truth caption $t_i$.


\section{Experiments and Results}
\label{sec:experiments}

{\noindent \textbf{Datasets.}} For training and evaluating RECAP, we use either Clotho \cite{drossos2020clotho}, AudioCaps \cite{kim2019audiocaps}, or a combination of both. Clotho has 3839/1045/1045 unique audios in train/dev/test splits, respectively, with five captions for each audio. AudioCaps has 49,838/495/975 with five captions except for the train set.
\vspace{1mm}


{\noindent \textbf{Baselines.}} We compare RECAP with six
competitive baselines that are taken from literature. Eren~\etal~\cite{eren2020audio} and Xu~\etal~\cite{xu2021investigating} train a Gated Recurrent Unit (GRU) for generating captions, conditioned on audio embeddings extracted from an audio encoder. Chen~\etal~\cite{chen2020audio} replaces the GRU with a transformer decoder, and Mei~\etal~\cite{mei2021audio} trains an entire encoder-decoder transformer architecture from scratch. Kim~\etal~\cite{kim2023prefix} and Gontier~\etal~\cite{gontier2021automated} use a pre-trained language model, where the former employs GPT-2, and the latter employs BART \cite{lewis-etal-2020-bart}.
\vspace{1mm}

{\noindent \textbf{Experimental Setup.}} To compare the performance of RECAP, we conduct experiments in three distinct setups: (1) We train and evaluate the model on the same dataset $\mathcal{D}$, (2) We train the model on a dataset $\mathcal{D}$ and evaluate the model on a different dataset $\hat{\mathcal{D}}$ (3) We train the model on a combination of both datasets and evaluate separately on individual datasets. For (1), the datastore $\mathcal{DS}$ consists of captions from either the training set of the source dataset $\mathcal{D}$ or a large curated datastore $\mathcal{DS}_{large}$. For (2), we use $\mathcal{DS}$ that has captions from either $\mathcal{D}$ ($\mathcal{DS}$), $\mathcal{DS}_{large}$ or from the other dataset. For (3), we either use $\mathcal{DS}$ that has captions from both datasets or use $\mathcal{DS}_{large}$. We list all the sources of $\mathcal{DS}_{large}$ with over 600,000+ text-only captions on our GitHub. This includes 100,000+ new weakly labeled captions for the AudioSet strong subset and three new captions for each sample in AudioCaps and Clotho. All these captions were generated using GPT-4 and manually corrected by one expert human annotator. For retrieval-based prompt creation, we use $k$=4 and retrieve only the top 4 captions from the datastore. It is worth noting that RECAP does not use any additional training or data augmentation tricks. For both AudioCaps and Clotho, we train using Adam optimizer with a learning rate of 5e$^{-5}$ for 100 epochs and a batch size of 32. We evaluate all our models on the metrics of BLEU, METEOR, ROUGE-L, CIDEr, SPICE, and SPIDEr.
\vspace{1mm}


{\noindent \textbf{Results.}} Table \ref{table:clotho} and Table \ref{table:audiocaps} compare the performance of RECAP against all our baselines evaluated on Clotho and AudioCaps, respectively. We train our models in different settings and evaluate them with different datastores. While RECAP shows decent margins of improvement in in-domain settings, RECAP outperforms all baselines by a significant margin in out-of-domain settings when an in-domain datastore is available. Without one, RECAP shows competitive performance with SOTA~\cite{kim2023prefix}. The presence of a larger data store ($\mathcal{DS}_{large}$) almost always improves performance. This opens possibilities to improve captioning performance by augmenting the datastore with diverse synthetically generated captions.

{\noindent \textbf{Results Analysis.}} Table \ref{tab:teaser-table} compares RECAP with Kim\etal~\cite{kim2023prefix} (SOTA) on compositional instances from Clotho (\textbf{1.}) and AudioCaps (\textbf{4.}) test set. While SOTA was able to caption only one audio event, due to conditioning on a prompt constructed from diverse retrieved captions, RECAP captures multiple. We also compared with a model trained on AudioCaps and inferred on a Clotho test instance with an audio event never seen during training (\textbf{2.}), and vice-versa (\textbf{3.}). By being conditioned on in-domain prompts, RECAP can caption these instances effectively.

\vspace{-1.5em}
\begin{table}[h!]
    \centering
    \caption{\small Comparing RECAP in 4 challenging settings.}
    \resizebox{\columnwidth}{!}{%
    \begin{tabular}{>{\footnotesize}l|>{\small}l}
    \toprule
        \multirow{4}{*} {\textbf{Ground Truth}}
         & 1: a engine roars in the background while pieces of metal are being dropped in.\\
         & 2:  a moving vehicle has some metal container in it clinging against each other. \\
        & 3: nature sounds with a frog croaking. \\
        & 4: a vehicle driving as a man and woman are talking and laughing. \\
        \midrule
        \multirow{4}{*} {\textbf{SOTA} \cite{kim2023prefix}}
         & 1: a bell is ringing and a bell rings. \\
         & 2: rain falling on a surface. \\
         & 3: people are talking and laughing with a man speaking in the background.\\
         & 4: a person is talking in the background. \\
        \midrule
        \multirow{4}{*}{\textbf{RECAP}}
        &\mycc 1: A person is using a chisel to cut wood and a car passes by.\\
        &\mycc 2:  Water splashes while a car drives by in the rain.\\
        &\mycc 3:  several vehicles move and a beep goes off.\\
        &\mycc 4:  an adult male is speaking, and a motor vehicle engine is running.\\
    \bottomrule
    \end{tabular}%
    }
    \label{tab:teaser-table}
\end{table}
\vspace{-1em}
\section{Conclusion and Future Work}
\label{sec:conclusion}

We present RECAP, a novel audio captioning system based on retrieval-augmented generation. While being competitive with state-of-the-art methods on benchmark datasets, RECAP outperforms SOTA by a huge margin on out-of-domain settings and shows unique capabilities of captioning novel audio events and compositional audios with two or more events. Additionally, RECAP is cheap to train and can exploit a replaceable text-caption-only datastore in a \textit{training-free} fashion to further push performance. As part of future work, we would like to explore advanced techniques for efficient retrieval and build better audio-text models.


\bibliographystyle{IEEEbib}
\bibliography{strings,refs}

\begin{thebibliography}{10}

\bibitem{wu2023large}
Yusong Wu, Ke~Chen, Tianyu Zhang, Yuchen Hui, Taylor Berg-Kirkpatrick, and
  Shlomo Dubnov,
\newblock ``Large-scale contrastive language-audio pretraining with feature
  fusion and keyword-to-caption augmentation,''
\newblock in {\em IEEE ICASSP 2023}.

\bibitem{kim2019audiocaps}
Chris~Dongjoo Kim, Byeongchang Kim, Hyunmin Lee, and Gunhee Kim,
\newblock ``Audiocaps: Generating captions for audios in the wild,''
\newblock in {\em ACL 2019}, pp. 119--132.

\bibitem{drossos2020clotho}
Konstantinos Drossos, Samuel Lipping, and Tuomas Virtanen,
\newblock ``Clotho: An audio captioning dataset,''
\newblock in {\em IEEE ICASSP 2020}, pp. 736--740.

\bibitem{koepke2022audio}
A~Sophia Koepke, Andreea-Maria Oncescu, Joao Henriques, Zeynep Akata, and
  Samuel Albanie,
\newblock ``Audio retrieval with natural language queries: A benchmark study,''
\newblock {\em IEEE Transactions on Multimedia}, 2022.

\bibitem{gontier2021automated}
F{\'e}lix Gontier, Romain Serizel, and Christophe Cerisara,
\newblock ``Automated audio captioning by fine-tuning bart with audioset
  tags,''
\newblock in {\em DCASE2021 Challenge}, 2021.

\bibitem{kim2023prefix}
Minkyu Kim, Kim Sung-Bin, and Tae-Hyun Oh,
\newblock ``Prefix tuning for automated audio captioning,''
\newblock in {\em IEEE ICASSP 2023}, pp. 1--5.

\bibitem{koizumi2020audio}
Yuma Koizumi, Yasunori Ohishi, Daisuke Niizumi, Daiki Takeuchi, and Masahiro
  Yasuda,
\newblock ``Audio captioning using pre-trained large-scale language model
  guided by audio-based similar caption retrieval,''
\newblock {\em arXiv preprint arXiv:2012.07331}, 2020.

\bibitem{lewis2020retrieval}
Patrick Lewis, Ethan Perez, Aleksandra Piktus, Fabio Petroni, Vladimir
  Karpukhin, Naman Goyal, Heinrich K{\"u}ttler, Mike Lewis, Wen-tau Yih, Tim
  Rockt{\"a}schel, et~al.,
\newblock ``Retrieval-augmented generation for knowledge-intensive nlp tasks,''
\newblock {\em NeurIPS 2020}, pp. 9459--9474.

\bibitem{eren2020audio}
Ay{\c{s}}eg{\"u}l~{\"O}zkaya Eren and Mustafa Sert,
\newblock ``Audio captioning based on combined audio and semantic embeddings,''
\newblock in {\em IEEE International Symposium on Multimedia}, 2020.

\bibitem{xu2021investigating}
Xuenan Xu, Heinrich Dinkel, Mengyue Wu, Zeyu Xie, and Kai Yu,
\newblock ``Investigating local and global information for automated audio
  captioning with transfer learning,''
\newblock in {\em ICASSP 2021}.

\bibitem{wu2023_t6a}
Wu~et~al.,
\newblock ``Beats-based audio captioning model with instructor embedding
  supervision and chatgpt mix-up,''
\newblock Tech. {R}ep., DCASE2023 Challenge.

\bibitem{kadlvcik2023whisper}
Marek Kadl{\v{c}}{\'\i}k, Adam H{\'a}jek, J{\"u}rgen Kieslich, and Rados{\l}aw
  Winiecki,
\newblock ``A whisper transformer for audio captioning trained with synthetic
  captions and transfer learning,''
\newblock {\em arXiv preprint arXiv:2305.09690}, 2023.

\bibitem{yan2023_t6a}
Haoran Sun, Zhiyong Yan, Yongqing Wang, Heinrich Dinkel, Junbo Zhang, and Yujun
  Wang,
\newblock ``Leveraging multi-task training and image retrieval with clap for
  audio captioning,''
\newblock Tech. {R}ep., DCASE2023 Challenge.

\bibitem{lee2023_t6a}
Jaeheon Sim, Eungbeom Kim, and Kyogu Lee,
\newblock ``Label-refined sequential training with noisy data for automated
  audio captioning,''
\newblock Tech. {R}ep., DCASE2023 Challenge.

\bibitem{berant2013semantic}
Jonathan Berant, Andrew Chou, Roy Frostig, and Percy Liang,
\newblock ``Semantic parsing on freebase from question-answer pairs,''
\newblock in {\em EMNLP 2013}, pp. 1533--1544.

\bibitem{kwiatkowski2019natural}
Tom Kwiatkowski, Jennimaria Palomaki, Olivia Redfield, Michael Collins, Ankur
  Parikh, Chris Alberti, Danielle Epstein, Illia Polosukhin, Jacob Devlin,
  Kenton Lee, et~al.,
\newblock ``Natural questions: a benchmark for question answering research,''
\newblock {\em TACL 2019}, pp. 453--466.

\bibitem{ramos2023retrieval}
Rita Ramos, Desmond Elliott, and Bruno Martins,
\newblock ``Retrieval-augmented image captioning,''
\newblock {\em arXiv preprint arXiv:2302.08268}, 2023.

\bibitem{Ramos_2023_CVPR}
Rita Ramos, Bruno Martins, Desmond Elliott, and Yova Kementchedjhieva,
\newblock ``Smallcap: Lightweight image captioning prompted with retrieval
  augmentation,''
\newblock in {\em Proceedings of the IEEE/CVF Conference on Computer Vision and
  Pattern Recognition (CVPR)}, June 2023, pp. 2840--2849.

\bibitem{mei2021audio}
Xinhao Mei, Xubo Liu, Qiushi Huang, Mark~D. Plumbley, and Wenwu Wang,
\newblock ``Audio captioning transformer,''
\newblock in {\em DCASE2021 Challenge}.

\bibitem{chen2020audio}
Kun Chen, Yusong Wu, Ziyue Wang, Xuan Zhang, Fudong Nian, Shengchen Li, and
  Xi~Shao,
\newblock ``Audio captioning based on transformer and pre-trained cnn.,''
\newblock in {\em DCASE}, 2020, pp. 21--25.

\bibitem{koh2022automated}
Andrew Koh, Xue Fuzhao, and Chng~Eng Siong,
\newblock ``Automated audio captioning using transfer learning and
  reconstruction latent space similarity regularization,''
\newblock in {\em ICASSP 2022}.

\bibitem{izacard2022few}
Izacard et~al.,
\newblock ``Few-shot learning with retrieval augmented language models,''
\newblock {\em arXiv preprint arXiv:2208.03299}, 2022.

\bibitem{li2022survey}
Huayang Li, Yixuan Su, Deng Cai, Yan Wang, and Lemao Liu,
\newblock ``A survey on retrieval-augmented text generation,''
\newblock {\em arXiv preprint arXiv:2202.01110}, 2022.

\bibitem{lewis-etal-2020-bart}
Mike Lewis, Yinhan Liu, Naman Goyal, Marjan Ghazvininejad, Abdelrahman Mohamed,
  Omer Levy, Veselin Stoyanov, and Luke Zettlemoyer,
\newblock ``{BART}: Denoising sequence-to-sequence pre-training for natural
  language generation, translation, and comprehension,''
\newblock in {\em ACL 2020}, pp. 7871--7880.

\end{thebibliography}

\end{document}